\documentstyle[12pt,aaspp4]{article}
\def\insertplot#1#2#3#4#5#6#7{
\vskip 10pt\nobreak\hbox to \hsize{\hss\dimen0=#3in\hbox to #6\dimen0{%
\dimen0=#2in\vbox to #6\dimen0{\vss
\special{ps: plotfile #1}
\special{ps::[end]
  PGPLOT restore
}
}\hss}\hss}\vskip 10pt}

\newcommand{\wvnum}{cm$^{-1}$}
\newcommand{\wv}{H$_2$O}

\lefthead{Wallace et al.}
\righthead{J--Band Classification Spectroscopy}
\begin{document}
\title{Near--Infrared Classification Spectroscopy:  J--Band Spectra of Fundamental MK Standards}

\author{Lloyd Wallace}
\affil{Kitt Peak National Observatory, NOAO\altaffilmark{1}, Tucson, AZ 85726}
 
\author{Michael R. Meyer\altaffilmark{2}}
\affil{ Steward Observatory, The University of Arizona, Tucson, AZ 85721--0065 \\
mmeyer@as.arizona.edu}

\author{Kenneth Hinkle}
\affil{Kitt Peak National Observatory, NOAO\altaffilmark{1}, Tucson, AZ 85726 \\
hinkle@noao.edu}

\centerline{and}
 
\author{Suzan Edwards}
\affil{Astronomy Department, Smith College, Northampton, Massachusetts 01063 }

\altaffiltext{1}{Operated by the Association of Universities for Research
in Astronomy, Inc.\ under cooperative agreement with the National Science
Foundation}

\altaffiltext{2}{Hubble Fellow}
 
\begin{abstract}

We present a catalog of J--band spectra for 88 fundamental MK standard
stars observed at a resolving power of $R \sim 3000$.  This contribution
serves as a companion atlas to the K--band spectra published by Wallace
and Hinkle (1997) and the H--band atlas of Meyer et al. (1998).  We
report data from 7400--9550 cm$^{-1}$ (1.05--1.34 $\mu$m) for stars of
spectral types O7--M6 and luminosity classes I--V as defined in the MK
system.  In reducing these data, special care has been taken to remove
time--variable telluric features of water vapor.   We identify atomic and molecular
indices which are both temperature and luminosity sensitive that aid in
the classification of stellar spectra in the J--band.  In addition to
being useful in the classification of late--type stars, the
J--band contains several features of interest in the study of
early--type stellar photospheres.  These data are available
electronically for anonymous FTP in addition to being served on the
World Wide Web.

\end {abstract}


\section {Introduction}

Over the last several years, there has been an explosion of interest in
near--infrared spectroscopy.  Improvements in infrared array detectors
have led to the construction of sensitive, high resolution
spectrographs for this wavelength regime (e.g. Hinkle et al. 1998;
McLean et al. 1999).  In addition, large area photometric sky surveys
such as 2MASS (Skrutskie et al. 1997) and DENIS (Epchtein, 1997) have
produced large target lists which require follow--up spectroscopy.  In
order to provide a comprehensive set of uniform high quality infrared
stellar spectra, we undertook a multi--wavelength survey of fundamental
MK standards with the KPNO Mayall 4 m telescope utilizing the Fourier
Transform Spectrograph.   Wallace and Hinkle (1997; hereafter WH97)
report observations of these stars in the K--band.  Meyer et al. (1998;
hereafter MEHS98) present the H--band data from our survey as well as
outline a classification scheme based on several atomic and molecular
indices.   This third paper in the series is devoted to presenting the
J--band spectra. 

Relatively little work has been done thus far in this wavelength range
compared to the H-- and K--bands.  The first atlas of stellar spectra
in the near--infrared was the pioneering work of Johnson and Mendez
(1970).  They present spectra of 32 stars from 1--4 $\mu$m with
resolving power varying from 300--1000 for stars of spectral type
A0--M7 as well as some carbon stars.  A review of early work in
infrared stellar spectroscopy is given by Merrill and Ridgeway (1979).
More recently, Kirkpatrick et al.  (1993) present spectra from 0.6--1.5
$\mu$m for a series of M dwarf standards from M2--M9.  They present a
detailed analysis of features useful for classification as well as
compare the spectra with model atmosphere calculations for cool stars.
Joyce et al. (1998a) has complemented this work with an atlas of
spectra obtained from 1.0--1.3 $\mu$m at $R = 1100$ for 103 evolved
stars of S--, C--, and M--type.  In this study, the dominant atomic and
molecular absorption features were identified and compared to
laboratory spectra.   For a review of more recent work concerning the
infrared spectra of stars in the K-- and H--bands, see WH97 and MEHS98
respectively.

Here we present a comprehensive atlas of stellar spectra from
7400--9500 cm$^{-1}$ (1.05--1.34 $\mu$m) to complement and extend the
work described above. In section 2, we describe the observations and
the data reduction.  In section 3, we present the spectra and detail
the identification of spectral features observed.  We explore stellar
classification based on J--band spectra in section 4, and discuss and 
summarize our results in section 5.

\section {Data Acquisition \&\ Reduction}

\subsection{Description of the Sample and Observations} 

Our sample of 88 stars is drawn from lists of fundamental MK spectral 
standards as follows; i) Morgan, Abt, \& Tabscott (1978) for stars O6--G0; 
ii) Keenan \& McNeil (1989) for stars G0--M6; and iii) Kirkpatrick, Henry,
\& McCarthy (1991) for late--type dwarfs K5--M3.  
The goal was to
observe bright well--established standard stars covering the full range
of spectral types (26 bins) and luminosity classes (three bins) in the
two--dimensional H--R diagram (78 bins total).   Secondary standards
were also added from compilations of Jaschek, Conde, \& de Sierra
(1964) and Henry, Kirkpatrick, and Simon (1994).  Due to the sensitivity 
limits of the FTS, we were unable to observe dwarf star standards with 
spectral types later than M3.  Our sample is nearly identical to that 
used in the H--band study of MEHS98 (82 stars in common).  For details
concerning the sample selection and stellar properties, see section II
in MEHS98.  In Table 1, we provide a journal of observations including 
the catalogue name of the source, its common name, the spectral type of 
the star, and the date of observation.  We also indicate which stars
analyzed are common to this sample and that of MEHS98, and which spectra
are used in the equivalent width analysis presented in section 4.  

The observations were obtained with the Mayall 4m telescope at Kitt Peak 
National Observatory in Arizona.  We employed the dual--output Fourier
Transform Spectrometer (FTS) described by Hall et al. (1979) to collect
spectra of our survey sample in the J-- and H--band simultaneously.  A
dichroic beam--splitter was used to separate the portions of the
incident flux longward and shortward of 1.5 $\mu$m and re--direct the
beams toward detectors equipped with the appropriate filters.

Each star was centered in a 3.8'' aperture and simultaneous
measurements of the sky were obtained through an identical aperture
offset 50'' in the east--west direction.  The interferogram was sampled
at 1 kHz in both the forward and backward scan directions as the path
difference was continuously varied from $\pm$0.75 cm yielding an
unapodized FWHM resolution of 0.8 cm$^{-1}$.  Data were obtained in a
beam--switching mode (A--B--B--A), alternating the source position
between the two input apertures.  The interferograms were co--added
keeping the opposite scan direction separate but combining data 
obtained in both apertures.  Because of the novel design of
the dual--input FTS, background emission from the night sky (obtained
from the offset aperture) is subtracted from the interferogram of the
star+sky spectrum in Fourier space as the data are collected.   The
resultant forward and backward scan pairs were transformed at KPNO with
output spectra of relative flux as a function of wavenumber ($\sigma$
in cm$^{-1}$).  Further description can be found
in MEHS98.

\subsection{Analysis of the Data}

Following the production of the transformed spectra, the next step in
the reduction is division by the spectrum of an essentially featureless
reference star taken with the same instrumental set-up and interpolated
to the same air mass.  This has the effect of removing the filter
response function and correcting for stable telluric absorbers in this 
spectral region (e.g. O$_2$).  Vega (A0 V) and Sirius (A1 V) were used as reference stars
after interpolation over photospheric absorption features due to H P$\beta$, H
P$\gamma$ and C I.  This is different from the technique used by MEHS98
where the time--constant portion of the atmospheric opacity was derived
from multiple observations of reference stars as a function of
airmass.  An example of removing the airmass--dependent portion of the
telluric absorption is shown in the center panel of Fig. 1 for HR1713
(B8 Ia).  This process
compensates well for O$_2$ but only partially for \wv\ because it can
vary over time and does not correlate strictly with airmass.  For this reason
MEHS98 separated the J-- and H--band portions of the spectra 
and concentrated on the H--band data in their paper.

We have taken the next step in correcting for the time variable \wv\
absorption, enabling a full analysis of these spectra.  To obtain the
\wv\ spectrum we ratioed two spectra of the same reference star, Vega,
obtained at similar air mass on different nights, but preferably in the
same observing run as the program star.  This required experimentation
with different spectra to obtain the largest \wv\ signal since some
spectra showed similar amounts of \wv.  This water spectrum was then
stretched logarithmically to match the residual \wv\ in the program
star spectrum remaining and divided out.  The result of this second step for
HR1713 is shown in the lower panel of Fig. 1.  The influence of \wv\ is
clearly reduced.

These corrections appear to be very sensitive to drift in instrumental
settings.  Because of this, it was essential to 
obtain frequent spectra of reference stars.  Weather problems
interfered with the 1992 October run resulting in inadequate reference spectra
and necessitating use of reference spectra from other runs.  This appears to 
have introduced a continuum distortion noticeable in some of the spectra. 
This distortion, illustrated in the lower panel of Fig. 1, 
is the result of low frequency variations between the derived \wv\ 
spectrum and the program object.  We suspect that the
character of the band-pass filter may have been temperature sensitive, 
though a definitive cause has not been established. 

The full list of 88 reduced spectra is given in Table 1, which includes
12 that have been observed on multiple runs and three 
for which observations on different nights of a single run have
been combined.  In general, Vega was used as the
reference star to achieve the telluric correction, the only exceptions
being; i) HR2197, HR2943, and HR2985 for which an interpolation between
Sirius and Vega was used; and ii) HR2827 for which Sirius alone was used.  
In Figures 2--5, we present the highest signal--to--noise ratio (SNR $>$ 25) spectra for each 
distinct spectral type in the survey:  18 supergiants, 19 giants, 
9 subgiants, and 19 dwarf stars. 
These spectra have been apodized with a gaussian filter for a resulting 
resolution of 2.7 \wvnum\ matching that of WH97 and MEHS98
(see WH97 for details concerning the apodization process). 
We utilize these 65 spectra in the analyses presented below. 

\section {Line Identifications}

In order to identify the dominant features observed, we
concentrated on four spectra of highest SNR spanning a wide range
of effective temperatures in luminosity classes I--II, III, and V.  These 
twelve spectra are shown at a resolution of 2.7 \wvnum\ \ 
with an expanded scale in Figures 6--8.  The spectra 
have also been shifted to zero velocity with respect to the 
laboratory frequencies of H and He lines (for early--type stars) 
and Al I lines (for late--type stars). 
The identifications are based on our previous studies of 
the solar photosphere (Livingston \&\ Wallace 1991; Wallace et al. 1993),
and sunspot umbrae (Wallace \&\ Livingston 1992; Wallace et al. 1998), 
as well as the high resolution Arcturus Atlas (Hinkle et al 1995).  A list of the
identified features is given in Table 2 along with references to
relevant laboratory data.  Many of these features are also labeled 
in Figures 6--8.  At the resolution of our data, some 
features in the spectra appear to be blends from multiple species.
For example, the feature at 8552 \wvnum\ 
appears to be due primarily to Fe I in stars of spectral type F--G.
However, for later spectral types K I absorption plays the dominant 
role in determining the feature strength.  In some cases, it was not
clear whether a particular species contributes to an observed spectral feature.   
For example, Sr II may play a role along with Mg II in determining the strength 
of the 9160 \wvnum\ feature.  However detailed modeling is required in order
to determine which species dominates.  Some of the spectra have high enough SNR 
that weaker features might also be identified but we stopped at what
seems to be a defensible level.

\section{Dependence on Temperature and Luminosity Class}

As illustrated in Figures 2--5 (\& 6--8), the J--band spectral region 
contains a large number of features
which are temperature and luminosity sensitive.  In the earliest type
stars, features of neutral hydrogen (Pa $\beta$ and Pa $\gamma$) and
helium dominate the spectra.  These features become weaker in stars of
intermediate spectral type (A, F, G), whose spectra are dominated by
lines of neutral metal species of lower ionization potential.  For
example, C I appears only in spectral types A through early G.  Si I
first appears in early F stars and fades by 
early M.  Al I, Mn I, Fe I, and Mg I behave similarly to Si I
but persist through the latest types observed.  Na I at 7880 cm$^{-1}$
first appears in early G and grows in strength toward later types
though it is relatively weak. 
K I, the lowest ionizational potential species observed in
these spectra, first appears in early K stars.  Even later type dwarfs
($>$ M5), which are not included in our survey, show very strong
features due to FeH in this wavelength regime (cf. Kirkpatrick et al.
1993; Jones et al. 1994).

The most striking luminosity sensitive feature in the spectra is the
0--0 band of CN at 9100 cm$^{-1}$.  This feature, observed in stars
later than G6 for luminosity class III stars is barely detected in
dwarf stars.  The 0--0 band of TiO at 9060 cm$^{-1}$ (obvious 
in the spectrum of the M6 III HR7886) may also play a role in
the latest type stars.  Ti I 
appears in the spectra of K--M giants but is much weaker in 
luminosity classes IV--V.  The Ca II features are observed in 
spectra of F--G supergiants, but are much weaker in stars of 
higher surface gravity.  These features are inferior to the 
first-- and second--overtone features of CO observed in the K-- \& 
H--bands, as well as the R-- \& I--band features of TiO as 
surface gravity diagnostics. 

In order to quantify the visual impressions made by the data, 
we have measured equivalent widths from these spectra for a number of prominent
features observed.  Following MEHS98, we have defined nine bandpasses and 
continuum regions with the intervals indicated in Table 3.  We 
measured the feature strengths with respect to the continuum levels 
by linear interpolation between two nearby regions.  Because the only available 
short wavenumber continuum region for CN is far from the feature
(``redward'' of a region of uncertain telluric correction 8650 to 9000 \wvnum )
we have estimated the equivalent width of CN using only a long wavenumber 
continuum reference.  The features measured for Al I, Mn I, Si I, Fe I, 
and C I are sums over two or more lines of the atom.  These nine equivalent
widths are reported in Table 4 for the 65 stars shown in Figures 2--5. 
Typical SNR for these indices as measured from multiple observations of 
the same stars are $>$ 10 except for Mn I (SNR $\sim$ 5) and CN (SNR $\sim$ 3). 
Adopting the spectral type to
effective temperature conversion as a function of luminosity class
given in MEHS98, we plot several of these equivalent width indices as a
function of T$_{eff}$ in Figure 9.  P $\beta$ shows
the expected behavior as a function of temperature with a peak in
strength near 10,000 K (A0).  The strength of the Mn I feature 
increases consistently with cooler temperatures in the spectra of 
the giant stars.  The Al I feature displays similar behavior in the 
dwarf star spectra, peaking in strength near 4000 K.  
Finally the CN index, although difficult to quantify because of
uncertainties in establishing accurate continuum levels, crudely indicates 
the dependence on luminosity class, being much stronger in the giants and
supergiants compared to the dwarfs.

In many cases, the strength of these features measured from stellar spectra
can be compared to our standards as a guide to
assigning spectral types.  However, it is often useful to examine
diagnostic line ratios which are not as sensitive to continuum excess
emission that can dilute straight equivalent widths.  In an attempt to
identify a two--dimensional classification plane similar to that
outlined in MEHS98, we have investigated several ratios from among the
features listed in Table 3.  For dwarf stars of spectral type F5--M2, 
the ratio of the Al I equivalent width to that of the Mg I feature
provides an estimate of the effective temperature.

\begin{equation}
T_{eff} (V) = 7100 \pm 390 - (2050 \pm  80) \frac{EW[Al I]}{EW[Mg I]}
\end{equation}

\noindent
Here EW is the equivalent width in cm$^{-1}$ for the indices
identified in Table 3 and listed in Table 4.  
For late--type giants G0--M6, the ration of the Mn I feature strength 
to that of Mg I gives a better temperature estimate than the  Al I feature. 

\begin{equation}
T_{eff} (III) = 6170 \pm 400 - (1860 \pm 90)\times \frac{EW[Mn I ]}{EW[Mg I]}
\end{equation}

Furthermore, the strength of
the molecular features of CN and TiO can be used to estimate the
surface gravity of the star, providing a rough estimate of the
luminosity class.  The temperature and
luminosity dependence of these diagnostics is illustrated in Figure 10.

Thus measurement of these line ratios provides an approximate 
classification of the spectral type and luminosity class for late--type
stars.  Based on the errors estimated from multiple observations of
several stars with $SNR > 25$, we conclude that crude spectral types can be
estimated within $\pm$ 3 subclasses (500 K) for late--type stars based on
these indices alone.

\section{Discussion and Summary}

Because stars earlier than M5 (3000 K)
have SEDs that peak at shorter wavelengths, features observed in the
spectra of these stars over the J--band are intrinsically weak.
However, there are several photospheric features present in this
wavelength regime which are diagnostic and observing them in the
J--band can help to penetrate large amounts of extinction.  Absorption
due to interstellar dust at the J--band is less than $1/3$ that
observed as visual wavelengths (e.g. Rieke and Lebofsky, 1985).  
For many objects it makes sense to
obtain spectra at even longer wavelengths such as the K--band (WH97).
However some targets surrounded by circumstellar material (such as
young stellar objects or evolved stars) exhibit continuum excess
emission due to thermal dust at temperatures as high at 1500--2000 K
(e.g. Meyer et al. 1997).  Because the emission from such hot dust
peaks at 1.5--2.0 $\mu$m, it may be advantageous to observe these
objects in the J--band.   Perhaps the greatest utility of J--band 
spectroscopy (not demonstrated in this paper) lies in the classification 
of very cool stars ($>$ M5), whose I-- and J--bands spectra are dominated 
by very broad atomic and molecular features (e.g. Kirkpatrick et al. 1993; 
Jones et al. 1994).

There are also technical reasons that make the J--band attractive:
it can be observed with a warm spectrograph without any
appreciable thermal background contributed by the optical train. 
However this requires a non-silicon based detector as standard CCD's 
are not efficient beyond 1.0 $\mu$m.  Joyce et al. (1998b) report 
detector--limited performance using a HgCdTe infrared detector 
in a dewar blocked with a cold J--filter in the focal plane of
an ambient temperature spectrograph while the H--band observations 
were still limited by thermal radiation from the spectrograph.
Further the intensity of the OH airglow emission, an important 
source of noise in low resolution 1--2.5 $\mu$m spectroscopic 
observations, is roughly twice as strong in the H--band compared 
with the J--band (Maihara et al. 1993).   However, in deciding 
whether or not to utilize the J--band spectral region for stellar 
classification, it must be remembered that telluric water vapor 
absorption can be difficult to treat over 
significant portions of the J--band spectral region. 

We present J--band spectra from 7400--9550 cm$^{-1}$ (1.05--1.34
$\mu$m) for 88 fundamental MK spectral standards.  The stars span a
range of spectral types from O5--M6 and cover luminosity class I--V.
Special care was taken to remove absorption due to time--variable
telluric \wv\ features which prevented analysis of these data in a
previously published paper (MEHS98).   We have identified
several features in the spectra which are temperature and luminosity
sensitive.  The absorption strengths of the nine most diagnostic 
features are tabulated over narrow--band indices.  We also define a
two--dimensional classification scheme based on these data, which makes
use of diagnostic line ratios rather than the equivalent widths of the
features alone.  Using this scheme, J--band spectra of stars spectral type 
G--M taken at $R \sim 3000$ with $SNR > 25$ can be classified within $\pm$ 3 
subclasses.  However, the very latest--type stars can be classified
with great precision at much lower spectral resolution.  We conclude
that the J--band contains many spectral features of interest for a wide
range of astrophysical studies and we hope that this contribution will
facilitate their use.

\section{Appendix A.  Electronic Availability of the Data} 

The 101 reduced spectra listed in Table 1 are available through the 
World Wide Web at http://www.noao.edu/archives.html.  The data are 
organized into directories according to year and month of observation.
Thus, the two spectra of HR1899 are in the directories 93mar and
94jan.  These spectra have been corrected for telluric absorption as
described above and scaled to unity in the region from 7900 to 8100
\wvnum, but not further processed.  The data will also be available through 
the Astronomical Data Center, NASA Goddard Space Flight Center, Code 
631, Greenbelt, MD 20771 (tel: 301--286--8310; fax: 301--286--1771; 
or via the internet at http://adc.gsfc.nasa.gov).  The raw FTS 
data are also available directly from NOAO (contact KHH for details). 

\acknowledgments

We thank Steve Strom for his expert assistance in guiding the 4m
telescope which enabled the collection of data presented in this
paper.  This research made use of the SIMBAD database operated by CDS
in Strasbourg, France as well as NASA's Astrophysics Data System
Abstract Service.  Support for M.R.M. was provided by NASA through
Hubble Fellowship grant HF--01098.01--97A awarded by the Space
Telescope Science Institute, which is operated by the Association of
Universities for Research in Astronomy, Inc. for NASA under contract
NAS 5--26555.  S.E. acknowledges support from the National Science 
Foundation's Faculty Award for Women Program. 



\clearpage

\begin{figure}
\caption{Illustration of the procedure for correcting the spectrum of HR1713
for telluric absorption.  The upper panel gives the observed spectrum, the
center panel gives the spectrum after correction to zero air mass, and the
lower panel gives the spectrum after specific correction for \wv.  The regions
appreciably affected by telluric absorption of \wv\ and O$_2$ are so indicated.
In the lower panel the large excursions between 8650 and 9000 \wvnum\ reflect
the uncertainty in the telluric correction in that region.
Note that the band pass is formed by \wv\ at the low frequency
end and by a cut-off filter at the high frequency end.  }
\label{redux}
\end{figure}

\begin{figure}
\caption{Representative spectra of luminosity classes I-II.  Only the more
prominent features are labeled.  Each spectrum is scaled to unity in the region
7900 to 8100 \wvnum\ and separated by a vertical shift of 0.5.  The region
8650 to 9000 \wvnum\ with the centered symbol $\bigoplus$\ is one in which the
\wv\ correction is not trustworthy and spurious features may arise.}
\label{rep_super}
\end{figure}

\begin{figure}
\caption{ Same as Fig. 2, but for luminosity class III.}
\label{rep_giant}
\end{figure}

\begin{figure}
\caption{ Same as Fig. 2, but for luminosity class IV.}
\label{rep_sub}
\end{figure}

\begin{figure}
\caption{ Same as Fig. 2, but for luminosity class V.}
\label{rep_dwarf}
\end{figure} 

\begin{figure}
\caption{
Fig. 6. - Identification of absorption features in selected spectra of
luminosity classes I-II.  Each spectrum is scaled to unity in the region
7900 to 8100 \wvnum\ and separated by a vertical shift of 0.5.  The region
8650 to 9000 \wvnum\ with the centered symbol $\bigoplus$\ is one in which the
\wv\ correction is not trustworthy and spurious features may arise.} 
\label{id_super}
\end{figure} 

\begin{figure}
\caption{ Same as Fig. 6, but for luminosity class III.}
\label{id_giant}
\end{figure}

\begin{figure}
\caption{ Same as Fig. 6, but for luminosity class V.} 
\label{id_dwarf}
\end{figure}

\begin{figure}
\caption{ Equivalent widths of some of the species presented in Table 4 plotted 
as a function of effective temperature for stars of high (right panel) 
and low (left panel) surface gravity.  Supergiants are plotted as open circles, 
giant stars as filled circles, subgiants as triangles, and dwarfs as stars. 
Typical SNR for the indices plotted are $>$ 10 except for 
Mn I (SNR $\sim 5$) and CN (SNR $\sim$ 3).}
\label{ew_teff}
\end{figure}

\begin{figure}
\caption{ Two--dimensional spectral classification for late--type 
giant and dwarf stars G--M using diagnostic lines ratios based on spectra 
with S/N $>$ 25.  Typical errors in the Al I and Mn I equivalent width ratios are 10 \%.}
\label{teff_lum}
\end{figure} 


\begin{deluxetable}{lccc lccc}
\tablewidth{0pt}
\tablecaption{Journal of Observations}
\tablenum{1}
\tablehead{
Star \tablenotemark{a} & Name & Sp. Type & Date \tablenotemark{b}  & Star \tablenotemark{a} & Name & Sp. Type & Date  \tablenotemark{b} 
}
\startdata
GL338A \tablenotemark{a} & ... & M0 V & 94/2/1                                    & HR4931 \tablenotemark{a} & 78 UMa & F2 V & 93/4/1               \nl
GL411  \tablenotemark{a} & ... & M2 V & 93/4/2  \tablenotemark{b}                 & HR4983 \tablenotemark{a} & $\beta$ Com & F9.5 V & 93/4/1        \nl
GL526  \tablenotemark{a,e} & ... & M1.5 V & 94/1/31                                 &   ...  &    ...      &   ...  & 94/1/30  \tablenotemark{b}     \nl
GL570A \tablenotemark{a} & ... & K4 V & 94/2/1   \tablenotemark{b}                & HR5017 \tablenotemark{a} & 20 CVn & F3 III & 93/4/2             \nl
GL725A \tablenotemark{a} & ... & M3 V & 93/4/2\&3   \tablenotemark{b}             & HR5072 \tablenotemark{a} & 70 Vir & G4 V & 94/2/1 \tablenotemark{b}                \nl
HR21   \tablenotemark{a} & $\beta$ Cas & F2 III-IV & 93/5/18  \tablenotemark{b}   & HR5191 \tablenotemark{a,d} & $\eta$ UMa & B3 V & 94/1/30  \tablenotemark{b}          \nl
HR152  \tablenotemark{a} & ... & K5-M0 III & 94/1/31                              &   ...  &    ...     &  000  & 92/10/16                           \nl
HR165  \tablenotemark{a} & $\delta$ And & K3 III & 94/1/31  \tablenotemark{b}     & HR5226 & 10 Dra & M3.5 III & 92/10/16  \tablenotemark{b}         \nl  
HR403  \tablenotemark{a} & $\delta$ Cas & A5 III-IV & 94/1/31                     & HR5235 \tablenotemark{a} & $\eta$ Boo & G0 IV & 93/1/26         \nl 
HR483  \tablenotemark{a} & ... & G1.5 V & 94/1/30                                 &   ...  &    ...     &  ...  & 94/1/30  \tablenotemark{b}       \nl                    
HR603  \tablenotemark{a} & $\gamma^1$ And & K3 -IIb & 94/1/31 \tablenotemark{b}   & HR5291 \tablenotemark{a} & $\alpha$ Dra & A0 III & 93/4/1  \tablenotemark{b}       \nl
HR995  \tablenotemark{a} & 59 Ari & G6 IV & 94/2/1   \tablenotemark{b}            & HR5409 \tablenotemark{a} & $\phi$ Vir & G2 IV & 94/1/30  \tablenotemark{b}  \nl 
HR1017 \tablenotemark{a} & $\alpha$ Per & F5 Ib & 93/4/1  \tablenotemark{b}       & HR5901 \tablenotemark{a} & $\kappa$ CrB & K1 IVa & 93/3/9       \nl 
HR1084 \tablenotemark{a} & $\epsilon$ Eri & K2 V & 94/1/31  \tablenotemark{b}     &   ...  &     ...      &  ...   & 93/4/3 \tablenotemark{b}      \nl 
HR1155 \tablenotemark{a} & BE Cam & M2 +IIab & 93/5/19  \tablenotemark{b}         & HR6014 \tablenotemark{a} & ... & K1.5 IV & 93/4/1               \nl 
HR1203 \tablenotemark{a} & $\zeta$ Per & B1 Ib & 93/4/1  \tablenotemark{b}        &   ...  & ... &  ...    & 94/1/31  \tablenotemark{b}            \nl
HR1279 \tablenotemark{a} & ... & F3 V & 94/1/31                                   & HR6242 \tablenotemark{a} & V636 Her & M4 +III-IIIa & 93/4/2  \tablenotemark{b}     \nl
HR1351 \tablenotemark{a} & 57 Tau & F0 IV & 94/1/30                               & HR6299 \tablenotemark{a} & $\kappa$ Oph & K2 III & 93/3/10      \nl
HR1412 \tablenotemark{a} & $\theta^2$ Tau & A7 III & 93/4/1                       & HR6322 & $\epsilon$ UMi & G5 III & 93/3/11  \tablenotemark{b}    \nl
HR1538 \tablenotemark{a,d} & 59 Eri & F6 V & 94/1/30  \tablenotemark{b}             & HR6406 \tablenotemark{a} & $\alpha^1$ Her & M5 Ib-II & 93/3/9  \tablenotemark{b}   \nl
HR1552 \tablenotemark{a} & $\pi^4$ Ori & B2+B2 III & 94/1/31  \tablenotemark{b}   & HR6498 \tablenotemark{a} & $\sigma$ Oph & K2 II & 93/3/9  \tablenotemark{b}        \nl
HR1577 & $\iota$ Aur & K3 II & 92/10/16                                           & HR6588 \tablenotemark{a} & $\iota$ Her & B3 IV & 93/4/3  \tablenotemark{b}         \nl
HR1713 \tablenotemark{a} & $\beta$ Ori & B8 Ia & 93/4/2  \tablenotemark{b}        & HR6623 \tablenotemark{a} & $\mu$ Her & G5 IV & 93/3/10  \tablenotemark{b}          \nl
HR1791 & $\beta$ Tau & B7 III & 92/10/16  \tablenotemark{b}                       & HR6705 \tablenotemark{a} & $\gamma$ Dra & K5 III & 93/1/26  \tablenotemark{b}    \nl       
HR1865 \tablenotemark{a} & $\alpha$ Lep & F0 Ib & 94/2/1  \tablenotemark{b}       &   ...  &     ...      &   ...  & 93/3/9       \nl
HR1899 \tablenotemark{a} & $\iota$ Ori & O9 III & 93/4/3 \tablenotemark{b}        & HR6713 \tablenotemark{a} & 93 Her & K0.5 IIb & 93/3/11  \tablenotemark{b}          \nl 
  ...  &    ...      &  ...   & 94/1/31                                           & HR7009 \tablenotemark{a} & XY Lyn & M4.5-5 +II & 93/3/9         \nl 
HR1903 \tablenotemark{a} & $\epsilon$ Ori & B0 Ia & 93/4/3 \tablenotemark{b}      & HR7063 \tablenotemark{a} & $\beta$ Sct & G4 IIa & 93/3/11  \tablenotemark{b}       \nl 
HR2061 \tablenotemark{a} & $\alpha$ Ori & M1-2 Ia-Iab & 93/4/2 \tablenotemark{b}  & HR7314 \tablenotemark{a} & $\theta$ Lyr & K0 +II & 93/3/9       \nl 
HR2197 & 6 Gem & M1-2 Ia-Iab & 92/10/16                                           &    ...  &     ...      &  ...   & 93/5/18      \nl 
HR2456 \tablenotemark{a} & 15 Mon & O7 V(e) & 93/4/1\&3                           & HR7328 \tablenotemark{a} & $\kappa$ Cyg & G9 III & 93/5/19  \tablenotemark{b}      \nl 
  ...  &   ...  &   ...   & 94/1/30  \tablenotemark{b}                            & HR7462 \tablenotemark{a} & $\sigma$ Dra & K0 V & 93/4/1  \tablenotemark{b}         \nl 
HR2706 \tablenotemark{a} & 48 Gem & F5 III-IV & 94/1/31                           & HR7479 \tablenotemark{a} & $\alpha$ Sge & G1 II & 93/3/11  \tablenotemark{b}       \nl 
HR2827 \tablenotemark{a} & $\eta$ CMa & B5 Ia & 94/2/1                            &   ...  &     ...      &  ...  & 93/4/3  \tablenotemark{b}      \nl 
HR2943 \tablenotemark{a} & $\alpha$ CMi & F5 IV-V & 92/10/16                      &   ...  &     ...      &  ...  & 93/5/19       \nl
  ...  &     ...      &   ...   & 94/1/31 \tablenotemark{b}                       & HR7602 \tablenotemark{a} & $\beta$ Aql & G8 IV  & 93/4/1  \nl  
HR2985 & $\kappa$ Gem & G8 III & 92/10/16                                         & HR7796 \tablenotemark{a} & $\gamma$ Cyg & F8 Ib & 93/3/11  \tablenotemark{b}       \nl
HR3275 & 31 Lyn & K4.5 III & 92/10/16                                             & HR7886 \tablenotemark{a} & EU Del & M6 III & 93/3/10  \tablenotemark{b}            \nl
HR3975 \tablenotemark{a} & $\eta$ Leo & A0 Ib & 94/2/1                            & HR7924 \tablenotemark{a} & $\alpha$ Cyg & A2 Ia & 93/3/11  \tablenotemark{b}       \nl
HR3982 \tablenotemark{a,d} & $\alpha$ Leo & B7 V & 94/1/30  \tablenotemark{b}       & HR7949 \tablenotemark{a} & $\epsilon$ Cyg & K0 -III & 93/5/19  \tablenotemark{b}   \nl
HR4031 \tablenotemark{a} & $\zeta$ Leo & F0 III & 94/1/30 \tablenotemark{b}       & HR7957 \tablenotemark{a} & $\eta$ Cep & K0 IV & 93/3/10 \tablenotemark{b}        \nl
HR4033 \tablenotemark{a} & $\lambda$ UMa & A2 IV & 93/4/1  \tablenotemark{b}      &   ...  &     ...    &  ...  & 93/5/19         \nl
HR4357 \tablenotemark{a,d} & $\delta$ Leo & A4 V & 94/1/30  \tablenotemark{b}       & HR8079 \tablenotemark{a} & $\xi$ Cyg & K4.5 Ib-II & 93/5/19     \nl
HR4374 \tablenotemark{a} & $\xi$ UMa B & G0 V & 93/4/2                            & HR8085 \tablenotemark{a} & 61 Cyg A & K5 V & 93/4/1  \tablenotemark{b}             \nl
HR4375 \tablenotemark{a} & $\xi$ UMa A & G0 V & 94/2/1  \tablenotemark{b}         & HR8086 \tablenotemark{a} & 61 Cyg B & K7 V & 93/4/2\&3  \tablenotemark{b}           \nl
HR4496 \tablenotemark{a} & 61 UMa & G8 V & 94/1/31  \tablenotemark{b}             & HR8089 \tablenotemark{a} & 63 Cyg & K4 Ib-IIa & 93/5/19  \tablenotemark{b}         \nl
HR4517 \tablenotemark{a} & $\nu$ Vir & M1 IIIab & 93/4/2  \tablenotemark{b}       & HR8232 \tablenotemark{a} & $\beta$ Aqr & G0 Ib & 93/5/18  \tablenotemark{b}        \nl
HR4534 \tablenotemark{a,d} & $\beta$ Leo & A3 V & 94/1/30  \tablenotemark{b}        & HR8317 \tablenotemark{a} & 11 Cep & K1 III & 93/4/3  \tablenotemark{b}             \nl
HR4716 \tablenotemark{a} & 5 CVn & G6 III & 94/1/30  \tablenotemark{b}            & HR8465 \tablenotemark{a} & $\zeta$ Cep & K1.5 Ib & 93/5/18  \tablenotemark{b}      \nl
HR4883 \tablenotemark{a} & 31 Com & G0 III & 93/4/3  \tablenotemark{b}            & HR8752 \tablenotemark{a} & V509 Cas & G4v0 \tablenotemark{c} & 93/5/18            \nl
HR4931 \tablenotemark{a} & 78 UMa & F2 V & 93/4/1   \tablenotemark{b}             &        &          &      &                    \nl 
\enddata
\tablenotetext{a}{Denotes spectra that also appear in the H--band atlas of MEHS98.}
\tablenotetext{b}{Denotes highest SNR data selected for indices listed in Table IV below.}
\tablenotetext{c}{Previously classified as G0Ia, but is intrinsically brighter.}
\tablenotetext{d}{Taken from Jaschek, Conde, and de Sierra (1964).}
\tablenotetext{e}{Taken from Henry, Kirkpatrick, and Simmons (1994).}
\end{deluxetable}

\begin{deluxetable}{ccccc}
\tablenum{2}
\tablewidth{0pt}
\tablecaption{Spectral Features Identified}
\tablehead{ $\sigma$\tablenotemark{a} & $\sigma$ & $\sigma$ & $\sigma$ & $\sigma$}
\startdata
H          & 9296.3     & Si I       & Ca II      & 7704.5     \nl
           & 9317.5     &            &            & 7750.0     \nl
7799.3     & 9336.8     & 7523.8     & 8366.1     &            \nl
9139.9     & 9350.9     & 7587.0     & 8444.4     & Fe I       \nl
           & 9357.0     & 8147.3     &            &            \nl
He I       &            & 8259.8     & Ti I       & 7762.0     \nl
           & Na I       & 8309.2     &            & 7903.8     \nl
7708.9     &            & 8336.9     & 7644.8     & 7910.0     \nl
7782.4     & 7884.8     & 8342.0     & 7683.2     & 7961.5     \nl
7819.7     &            & 9073.6     & 7781.8     & 8349.8     \nl
8352.6     & Mg I       & 9105.6     & 7791.2     & 8412.8     \nl
9160.9     &            & 9184.2     & 7797.2     & 8484.3     \nl
9231.0     & 8273.4     & 9197.5     & 7848.1     & 8552.0     \nl
           & 8452.1     & 9219.3     & 7889.8     & 8590.0     \nl
C I        & 9117.1     & 9233.6     & 7934.2     & 8612.7     \nl
           & 9247.2     & 9268.0     & 8366.2     & 8623.1     \nl
7925.5     &            & 9300.3     & 8406.1     &            \nl
7933.4     & Mg II      & 9377.5     & 8474.3     & Sr II      \nl
7945.9     &            & 9428.3     & 9278.3     &            \nl
7953.9     & 9128.4     & 9444.6     & 9314.6     & 9159.3     \nl
7958.3     & 9159.8     &            & 9320.3     &            \nl
7966.3     &            & K I        & 9376.9     & CN Red     \nl
8404.1     & Al I       &            & 9445.1     &            \nl
8427.3     &            & 7983.7     &            & 9118.7     \nl
8437.4     & 7602.0     & 8041.4     & Mn I       & 9150.4     \nl
8505.4     & 7617.9     & 8493.0     &            &            \nl
8509.6     &            & 8551.8     & 7506.0     & TiO $\phi$ \nl
8574.4     &            &            & 7520.3     &            \nl
           &            &            &            & 9062.1     \nl

\enddata
\tablenotetext{a}{Units cm$^{-1}$.
Units cm$^{-1}$.  References for frequencies:  H
Garcia \& Mack (1965); He I Martin (1960); C I Johansson \& Litz\'{e}n
(1964) and Johansson (1965); Na I and K I Risberg (1956); Mg I Risberg
(1964); Mg II Risberg (1955); Al I Bi\'{e}mont \& Brault (1987); Si I
Litz\'{e}n (1964); Ca II Edl\'{e}n \& Risberg (1956); Ti I Forsberg
(1991); Mn I Taklif (1990); Fe I Litz\'{e}n \& Verg\`{e}s (1976); Sr II
Newsom et al.  (1973); CN Red Davis \& Phillips (1963; The CN Red
features are the R$_1$ and R$_2$ heads of the 0-0 band); and TiO $\phi$
Phillips (1973; the TiO $\phi$ feature is the R head of the 0-0 band).}
\end{deluxetable}

\begin{deluxetable}{cccc}
\tablewidth{0pt}
\tablecaption{Defining Bandwidths.\tablenotemark{a}}
\tablenum{3}
\tablehead{
Feature & Feature & First Continuum & Second Continuum \nl
 & Limits, $\sigma$ & Level Limits, $\sigma$ & Level Limits, $\sigma$
}
\startdata
Al I       & 7146 - 7174 & 7087 - 7120 & 7203 - 7226 \nl
Mn I       & 7695 - 7756 & 7652 - 7677 & 7765 - 7777 \nl
H P$\beta$ & 7786 - 7807 & 7710 - 7736 & 7827 - 7862 \nl
He I       & 7807 - 7825 & 7710 - 7736 & 7827 - 7862 \nl
Si I       & 8302 - 8346 & 8197 - 8231 & 8372 - 8395 \nl
Mg I       & 8447 - 8455 & 8372 - 8395 & 8515 - 8538 \nl
Fe I       & 8581 - 8633 & 8515 - 8538 & 8633 - 8659 \nl
CN         & 9003 - 9155 & .....       & 9383 - 9422 \nl
C I        & 9331 - 9363 & 9275 - 9288 & 9454 - 9490 \nl
\enddata
\tablenotetext{a}{Units cm$^{-1}$.}
\end{deluxetable}

\begin{deluxetable}{cccccccccccc}
\tablewidth{0pt}
\tablecaption{Equivalent Widths of Spectral Indices \tablenotemark{a}} 
\tablenum{4}
\tablehead{
Source & T$_{eff}$ & Class & Al I & Mn I & P $\beta$ & He I & Si I & Mg I & Fe I & CN \& TiO & C I }
\startdata
HR1903  &  26001 & Ia  & 0.07  & 0.50  & 1.51  &  1.00  & -0.21  & -0.07 &  -0.16 & -0.68  & 0.50 \nl 
HR1203  &  20701 & Ib  & -0.31  & 0.73  & 1.74  &  0.81  & 0.09  & -0.04 &  0.51 & -1.70  & 0.42 \nl 
HR1713  &  11194 & Ia  & -0.01  & 0.40  & 1.36  &  0.07  & -0.12  & -0.09 &  -0.04 & -1.09  & 0.57 \nl 
HR7924  &  9078  & Ia & 0.09  & 0.26  & 1.09  &  -0.23  & 0.03  & -0.02 &  0.39 & 0.19  & 1.24 \nl 
HR1865  &  7691  & Ib & 0.20  & 0.16  & 2.48  &  0.05  & 0.70  & 0.22 &  0.84 & -0.11  & 1.34 \nl 
HR1017  &  6637  & Ib & 0.48  & -0.02  & 1.98  &  -0.02  & 1.22  & 0.28 &  1.29 & -0.15  & 2.97 \nl 
HR7796  &  6095  & Ib & 0.54  & 0.16  & 1.47  &  -0.13  & 1.43  & 0.40 &  1.65 & -0.12  & 3.50 \nl 
HR8232  &  5508  & Ib & 0.54  & 0.49  & 1.27  &  -0.02  & 1.40  & 0.49 &  1.96 & 0.71  & 2.34 \nl 
HR7479  &  5333  & II & 0.50  & 0.30  & 1.17  &  -0.18  & 1.01  & 0.35 &  0.79 & 2.28  & 1.45 \nl 
HR7063  &  4897  & IIa & 0.37  & 0.43  & 0.83  &  -0.25  & 1.56  & 0.55 &  1.25 & 6.09  & 1.33 \nl 
HR6713  &  4375  & IIb & 0.45  & 0.84  & 0.60  &  -0.21  & 1.41  & 0.57 &  1.22 & 4.99  & 1.40 \nl 
HR8465  &  4295  & Ib & 0.81  & 1.15  & 0.98  &  -0.09  & 1.79  & 0.86 &  1.77 & 9.34  & -0.06 \nl 
HR6498  &  4255  & II & 0.59  & 1.10  & 0.60  &  -0.27  & 1.25  & 0.71 &  1.25 & 6.63  & 0.26 \nl 
HR603  &   4130  & IIb & 0.50  & 1.44  & 0.78  &  -0.23  & 1.19  & 0.72 &  1.68 & 5.18  & 0.84 \nl 
HR8089  &  3990  & Ib-IIa & 0.93  & 1.62  & 0.89  &  -0.19  & 0.97  & 0.93 &  0.53 & 6.26  & -0.07 \nl 
HR2061  &  3548  & Ia-Iab   & 0.85  & 2.23  & 1.14  &  -0.34  & 2.05  & 0.87 &  2.21 & 8.17  & 0.37 \nl 
HR1155  &  3451  & IIab    & 0.97  & 1.82  & 0.83  &  -0.22  & 1.21  & 0.79 &  1.73 & 3.61  & -0.01 \nl 
HR6406  &  2798  & Ib-II   & 0.84  & 1.78  & 0.61  &  -0.39  & 1.21  & 0.68 &  1.16 & 3.26  & 0.22 \nl \hline 

HR1899  &  31988  & III   & 0.10  & 0.17  & 1.82  &  0.76  & -0.10  & -0.01 &  0.29 & 0.22  & 0.56 \nl 
HR1552  &  20276  & III   & 0.14  & 0.73  & 2.51  &  1.18  & -0.14  & -0.20 &  -0.28 & 0.47  & -1.79 \nl 
HR1791  &  13212  & III   & -0.38  & -0.42  & 2.91  &  0.43  & 0.41  & -0.08 &  -0.66 & -1.77  & 1.83 \nl 
HR5291  &  10092  & III   & -0.20  & -0.52  & 3.99  &  0.88  & 0.04  & 0.05 &  0.18 & -1.46  & 0.23 \nl 
HR403  &    8090  & III-IV   & 0.02  & -0.49  & 3.84  &  0.86  & 0.69  & 0.17 &  0.89 & -1.17  & 2.31 \nl 
HR4031  &   7144  & III   & 0.23  & -0.20  & 2.67  &  0.27  & 0.75  & 0.20 &  0.90 & 0.05  & 2.25 \nl 
HR21  &     6870  & III-IV   & 0.47  & -0.39  & 2.81  &  0.47  & 0.81  & 0.33 &  0.57 & 0.70  & 1.41 \nl 
HR4883  &   5915  & III   & 0.47  & 0.06  & 1.42  &  -0.03  & 1.12  & 0.34 &  0.68 & 0.48  & 2.03 \nl 
HR6322  &   5116  & III   & 0.33  & 0.66  & 0.76  &  -0.11  & 1.13  & 0.60 &  1.46 & 1.55  & 0.67 \nl 
HR4716  &   5046  & III   & 0.39  & 0.21  & 0.51  &  -0.29  & 1.33  & 0.67 &  1.12 & 1.73  & 0.83 \nl 
HR7328  &   4886  & III   & 0.78  & 0.47  & 0.78  &  -0.05  & 0.84  & 0.70 &  0.58 & 3.79  & -0.04 \nl 
HR7949  &   4808  & III   & 0.60  & 0.55  & 0.66  &  -0.06  & 0.63  & 0.65 &  1.12 & 3.53  & 0.29 \nl 
HR8317  &   4709  & III   & 0.95  & 0.64  & 0.60  &  -0.26  & 1.49  & 0.69 &  0.47 & 6.43  & 0.90 \nl 
HR165  &    4315  & III   & 0.73  & 1.22  & 0.77  &  -0.22  & 1.54  & 0.80 &  1.45 & 4.21  & 0.93 \nl 
HR6705  &   3990  & III   & 0.69  & 1.43  & 0.57  &  -0.33  & 1.34  & 0.82 &  1.32 & 4.75  & 0.21 \nl 
HR4517  &   3775  & IIIab   & 0.48  & 0.97  & 0.69  &  -0.18  & 0.87  & 0.69 &  0.98 & 1.79  & 0.59 \nl 
HR5226  &   3614  & III   & 0.80  & 1.27  & 0.57  &  -0.21  & 1.33  & 0.54 &  1.24 & -1.33  & -0.56 \nl 
HR6242  &   3556  & III-IIIa   & 0.92  & 1.57  & 0.63  &  -0.41  & 0.92  & 0.72 &  1.85 & 2.08  & 0.24 \nl 
HR7886  &   3250  & III   & 0.89  & 1.54  & 0.49  &  -0.47  & 0.88  & 0.66 &  1.05 & 2.05  & 0.53 \nl \hline 
HR6588  &  19010  & IV   & 0.21  & 0.32  & 3.05  &  1.07  & 0.21  & -0.03 &  -0.83 & -0.17  & 0.02 \nl 
HR4033  &   8810  & IV   & 0.10  & -0.38  & 4.37  &  1.01  & 0.01  & -0.03 &  1.07 & -0.64  & 1.47 \nl 
HR5235  &   5929  & IV   & 0.59  & 0.02  & 1.64  &  0.04  & 1.35  & 0.54 &  1.24 & 0.75  & 1.46 \nl 
HR5409  &   5834  & IV   & 0.53  & 0.34  & 1.11  &  0.00  & 1.35  & 0.39 &  1.15 & 1.71  & 2.01 \nl 
HR6623  &   5675  & IV   & 0.68  & 0.24  & 1.03  &  -0.16  & 1.60  & 0.73 &  1.25 & 2.63  & 0.62 \nl 
HR995  &    5623  & IV   & 0.46  & 1.07  & 0.49  &  -0.22  & 0.89  & 0.65 &  1.44 & 2.28  & 0.37 \nl 
HR7957  &   5236  & IV   & 0.75  & 0.82  & 0.45  &  -0.38  & 1.04  & 0.59 &  0.68 & 1.82  & 0.43 \nl 
HR5901  &   5128  & IVa   & 0.67  & 0.55  & 0.74  &  -0.20  & 1.33  & 0.80 &  0.70 & 3.85  & 1.02 \nl 
HR6014  &   5069  & IV   & 0.59  & 0.48  & 0.92  &  -0.08  & 1.27  & 0.61 &  1.70 & 2.83  & 1.48 \nl \hline 
HR2456  &  38018  & V(e)   & -0.45  & 0.53  & 1.44  &  0.61  & -0.63  & 0.06 &  0.17 & 0.70  & 0.00 \nl 
HR5191  &  19010  & V   & -0.14  & 0.04  & 3.03  &  0.79  & -0.05  & -0.06 &  -0.01 & -0.61  & -0.22\nl 
HR3982  &  13001  & V   & 0.23  & -0.53  & 3.07  &  0.59  & -0.13  & -0.05 &  -0.10 & -0.64  & 0.26 \nl 
HR4534  &   8590  & V   & 0.09  & -0.70  & 4.07  &  1.13  & 0.44  & 0.22 &  0.55 & -1.51  & 1.27 \nl 
HR4357  &   8375  & V   & 0.08  & -0.82  & 3.97  &  1.00  & 0.38  & 0.20 &  0.46 & -1.12  & 1.69 \nl 
HR4931  &   6745  & V   & 0.05  & -0.04  & 2.79  &  0.54  & 1.20  & 0.33 &  0.85 & -0.10  & 2.12 \nl 
HR2943  &   6531  & IV-V   & 0.17  & 0.12  & 2.47  &  0.35  & 0.70  & 0.24 &  0.92 & -0.24  & 1.49 \nl 
HR1538  &   6382  & V   & 0.10  & 0.02  & 2.47  &  0.26  & 0.77  & 0.29 &  1.37 & -0.75  & 1.86 \nl 
HR4983  &   5929  & V   & 0.34  & 0.15  & 1.52  &  -0.01  & 0.99  & 0.42 &  1.29 & 0.98  & 0.86 \nl 
HR4375  &   6081  & V   & 0.41  & 0.05  & 1.18  &  -0.00  & 1.14  & 0.60 &  0.91 & 1.40  & 0.34 \nl 
HR5072  &   5741  & V   & 0.74  & 0.43  & 0.91  &  0.03  & 1.21  & 0.69 &  0.91 & 1.50  & -0.08 \nl 
HR4496  &   5432  & V   & 0.50  & 0.77  & 1.10  &  -0.03  & 2.15  & 0.88 &  0.82 & 1.01  & 1.70 \nl 
HR7462  &   5236  & V   & 0.78  & 0.64  & 0.71  &  -0.26  & 1.55  & 0.81 &  0.58 & 0.91  & 1.00 \nl 
HR1084  &   5011  & V   & 0.84  & 0.61  & 0.73  &  -0.20  & 1.66  & 0.90 &  0.55 & 1.33  & 1.32 \nl 
Gl570A  &   4560  & V   & 1.29  & 0.39  & 0.59  &  -0.09  & 1.56  & 1.03 &  0.93 & 1.53  & -0.06 \nl 
HR8085  &   4335  & V   & 1.49  & 0.79  & 0.15  &  -0.31  & 1.04  & 1.22 &  0.64 & 0.03  & 0.72 \nl 
HR8086  &   4036  & V   & 1.62  & 0.72  & 0.47  &  -0.25  & 0.93  & 0.97 &  0.48 & -1.47  & 0.79 \nl 
Gl411  &    3531  & V   & 1.46  & 0.31  & 0.42  &  -0.13  & 0.33  & 0.68 &  1.15 & -0.19  & 0.94 \nl 
Gl725A  &   3380  & V   & 1.09  & 0.40  & 0.34  &  -0.34  & 0.99  & 0.68 &  0.63 & -0.86  & -0.34 \nl 
\enddata
\tablenotetext{a}{Units cm$^{-1}$.}
\end{deluxetable}
\end{document}